\documentstyle[preprint,tighten,epsfig,aps]{revtex}
\begin{document}
\baselineskip 1.05 \baselineskip

\setlength{\parindent}{1cm}
\def\A{{\cal{A}}}
\def\B{{\cal{B}}}
\def\C{{\cal{C}}}
\def\D{{\cal{D}}}
\def\H{{\cal{H}}}
\def\L{{\cal{L}}}
\def\M{{\cal{M}}}
\def\O{{\cal{O}}}
\def\PP{{\cal{P}}}
\def\Q{{\cal{Q}}}
\def\R{{\cal{R}}}

\def\bc{\begin{center}}
\def\ec{\end{center}}
\def\beq{\begin{equation}}
\def\eeq{\end{equation}}
\def\bea{\begin{eqnarray}}
\def\eea{\end{eqnarray}}
\def\bit{\begin{itemize}}
\def\eit{\end{itemize}}
\def\ben{\begin{enumerate}}
\def\een{\end{enumerate}}
\def\ba{\begin{array}}
\def\ea{\end{array}}
\def\bc{\begin{center}}
\def\ec{\end{center}}

\def\nl{\nonumber\\}

\setlength{\parskip}{1.5ex plus 0.5ex minus 0.5ex}

\def\beqar{\begin{eqnarray}}
\def\ra{\rightarrow}
\def\eeqar{\end{eqnarray}}
\def\be{\begin{eqnarray}}
\def\ee{\end{eqnarray}}
\def\beqast{\begin{eqnarray*}}
\def\eeqast{\end{eqnarray*}}
\def\be{\begin{enumerate}}
\def\ee{\end{enumerate}}
\def\lag{\langle}
\def\rag{\rangle}
\def\fnote#1#2{\begingroup\def\thefootnote{#1}\footnote{#2}
\addtocounter{footnote}{-1}\endgroup}
\def\beq{\begin{equation}}
\def\eeq{\end{equation}}
\def\haf{\frac{1}{2}}
\def\pa{\partial}
\def\plb#1#2#3#4{#1, Phys. Lett. {\bf B#2}, #3 (#4)}
\def\npb#1#2#3#4{#1, Nucl. Phys. {\bf B#2}, #3 (#4)}
\def\prd#1#2#3#4{#1, Phys. Rev. {\bf D#2}, #3 (#4)}
\def\prl#1#2#3#4{#1, Phys. Rev. Lett. {\bf #2}, #3 (#4)}
\def\mpl#1#2#3#4{#1, Mod. Phys. Lett. {\bf A#2}, #3 (#4)}
\def\rep#1#2#3#4{#1, Phys. Rep. {\bf #2}, #3 (#4)}
\def\llp#1#2{\lambda_{#1}\lambda'_{#2}}
\def\lplp#1#2{\lambda'_{#1}\lambda'_{#2}}
\def\slash#1{#1\!\!\!\!\!/}
\def\rpv{\slash{R_p}~}
\def\hs{\hat s}
\def\ca{{\cal A}}
\def\cb{{\cal B}}
\def\cc{{\cal C}}

%%%%%%%%%%%%%%%%%%%%%%%%%%%%%%%%%%%%%%%%%%%%%%%%%%%%%%%%%%%%%%%%%%%%%%%%%%

\draft
\preprint{
\begin{tabular}{r}
KAIST-TH 98/21
\\
hep-ph/yymmddd
\end{tabular}
}
\title{
Constraints on the R-parity Violating
Couplings from
$B^\pm \ra l^\pm \nu$ Decays 
}
\author{
Seungwon Baek
\thanks{E-mail: swbaek@muon.kaist.ac.kr}
and
Yeong Gyun Kim
\thanks{E-mail: ygkim@muon.kaist.ac.kr}
}
\address{
Department of Physics, Korea Advanced Institute of Science and
Technology \\
Taejon 305-701, Korea \\
}

\maketitle

\begin{abstract}
We derive the upper bounds on certain products of R-parity- and
lepton-flavor-violating couplings from
$B^\pm \ra l^\pm \nu$ decays.
These modes of $B$-meson decays
can constrain the product combinations of the couplings with
one or more heavy generation indices
which are comparable with or stronger than the present bounds.
And we investigate the possible effects of R-parity violating 
interactions on $B_c \ra l \nu$ decays.
These decay modes can be largely affected by R-parity violation. 
\end{abstract}


%\section{introduction}
%{\bf 1.}
%In the Standard Model (SM), there are no couplings which violate
%the baryon number $B$ and the lepton number $L$.
In supersymmetric extensions of the standard model, there are
gauge invariant interactions which violate 
the baryon number ($B$) and the lepton number ($L$)
in general. To prevent occurrings of these $B$- and $L$-violating
interactions in supersymmetric extensions of the standard model,
the additional global symmetry is required.
This requirement leads to the consideration of
the so called R-parity($R_p$).
Even though the requirement of $R_p$ conservation makes a theory
consistent with
present experimental searches, there is no good theoretical justification
for this requirement. Therefore
the models with explicit $R_p$-violation have been considered by
many authors \cite{ago}.

In the MSSM,
the most general $R_p$-violating superpotential is given by
\beq
W_{R\!\!\!\!/_p}=\lambda_{ijk}L_iL_jE_k^c+\lambda_{ijk}'L_iQ_jD_k^c+
\lambda_{ijk}''U_i^cD_j^cD_k^c.
\eeq
Here $i,j,k$ are generation indices and we assume that possible bilinear terms
$\mu_i L_i H_2$ can be rotated away.
$L_i$ and $Q_i$ are the $SU(2)$-doublet lepton and quark superfields and
$E_i^c,U_i^c,D_i^c$ are the singlet superfields respectively.
$\lambda_{ijk}$ and
$\lambda_{ijk}''$ are antisymmetric under the interchange of the first two and
the last two generation indices respectively; $\lambda_{ijk}=-\lambda_{jik}$ and
$\lambda_{ijk}''=-\lambda_{ikj}''$. So the number of couplings is 45 (9 of the
$\lambda$ type, 27 of the $\lambda'$ type and 9 of the $\lambda''$ type).
Among these 45 couplings, 36 couplings are related with the lepton
flavor violation.

%\section{present bounds}
There are upper bounds on a {\it single} $R_p$ violating coupling from
several different sources \cite{han,beta,numass,agagra}.
Among these, upper bounds from
neutrinoless double beta decay \cite{beta}, $\nu$ mass \cite{numass} and
$K^+,t-$quark decays \cite{agagra} are strong.
Neutrinoless double beta decay gives
$\lambda'_{111}<3.5\times10^{-4}$.
The bounds from $\nu$ mass are $\lambda_{133}<3\times10^{-3}$
and $\lambda'_{133}<7\times10^{-4}$.
From $K^+$-meson decays one obtain $\lambda'_{ijk}<0.012$ for $j=1$ and 2.
These bounds from $K^+$-meson decays are basis-dependent \cite{agagra,bha}.
Here all masses of scalar partners which mediate the processes are assumed
to be 100 GeV.
Extensive reviews of the updated limits on a single $R_p$-violating
coupling can be found in \cite{bha}.

There are more stringent bounds on some products of
the $R_p$-violating couplings from the mixings of the
neutral $K$- and $B$- mesons and rare leptonic decays of
the $K_L$-meson, the muon and the tau \cite{choroy}, $B^0$ decays
into two charged leptons \cite{Lee}, $b\bar{b}$ productions at LEP \cite{Feng}
and muon(ium) conversion, and $\tau$ and $\pi^0$ decays \cite{Ko},
semileptonic decays of $B$ mesons \cite{jang}, 
$B \rightarrow X_s l^+_i l^-_j$ decays \cite{kim}.

In this paper, we derive
the upper bounds on certain products of $R_p$ and
lepton flavor violating couplings from $B^\pm \ra l^\pm \nu$ decays
in the minimal supersymmetric
standard model (MSSM) with explicit $R_p$ violation.
These modes of $B$-meson decays
can constrain the product combinations of the couplings with
one or more heavy generation indices.
Here, we assume that the baryon number 
violating couplings $\lambda''$'s
vanish in order to avoid too fast proton decays.  Especially in the models with
a very light gravitino ($G$) or axino ($\tilde{a}$), $\lambda''$ have to be
very small independently of $\lambda'$ 
from the proton decay $p\rightarrow K^+G~({\rm or}~K^+\tilde{a})$ 
; $\lambda''_{112}<10^{-15}$ \cite{Choi}. 
One can construct a grand unified model which has only lepton number
non-conserving trilinear operators in the low energy superpotential when
$R_p$ is broken only by bilinear terms of the form $L_i H_2$
\cite{hasu}. And usually it may be very difficult to discern signals of
$B$-violating interactions above QCD backgrounds \cite{numass}.

%\vspace{1.0cm}

%\section{four-fermion interactions}
%{\bf 2.}
In the MSSM with $R_p$, the terms in
the effective lagrangian relevant for the leptonic $B$-meson decays are
%\cite{GrLi}
\beq
{\cal L}^{eff}(b \bar{q} \rightarrow  e_l \bar{\nu}_l)=
-V_{qb}\frac{4G_F}{\sqrt{2}}\left[
(\bar{q}\gamma^{\mu}P_Lb)(\bar{e}_l\gamma_{\mu}P_L\nu_l)-
R_l(\bar{q}P_Rb)(\bar{e}_lP_L\nu_l)\right],
\eeq
where
$P_{R,L}=\frac{1}{2}(1\pm\gamma_5)$,
%\beq
$R_l=r^2 m_{e_l} m_b^Y$, 
$r=\frac{\tan\beta}{m_{H^{\pm}}}$.
%\eeq
An upper index $Y$ denotes the running quark mass,
$\tan\beta$ is the ratio of the vacuum expectation values of the neutral
Higgs fields
and $m_{H^{\pm}}$ is the mass of the charged Higgs fields.
The first term in Eq. (2) gives the standard model (SM)
contribution and the second one
gives that of the charged Higgs scalars. Neglecting the masses of
the electron ($l=1$) and the muon ($l=2$), the contribution
of the charged Higgs scalars is zero. The contribution of the charged Higgs
scalars is not vanishing only when $l=3$ ;
$b \bar{q} \rightarrow \tau~\bar{\nu}_{\tau}$.
We neglect a term
proportional to $m_c^Y$ for $q=c$
since the term is suppressed by the mass ratio
$m_c^Y/m_b^Y$ and does not have the possibly large $\tan^2\beta$ factor.

In the MSSM without $R_p$, the exchange of the sleptons and the
squarks leads to the additional four-fermion
interactions which are
relevant for the leptonic decays of $B$-meson.
Considering the fact that the CKM matrix $V$ is not an identity matrix,
the $\lambda'$ terms of the Eq. (1) are reexpressed in terms of the
the fermion mass eigenstates as follow
\beq
W_{\lambda'}=\lambda'_{ijk}\left(N_iD_j-
\sum_{p}V^{\dagger}_{jp}E_iU_p\right)D^c_k,
\eeq
where $N_i$, $E_i$, $U_i$ and $D_i$ are the superfields with neutrinos,
charged leptons, up- and down-type-quarks and
$\lambda'$ have been redefined to absorb some field rotation effects.
From Eq. (1) and Eq. (3) we obtain the effective interactions which are
relevant for the leptonic decays of $B$-meson as follows
\beq
{\cal L}^{eff}_{\slash{R_p}}(b \bar{q} \rightarrow  e_l \bar{\nu}_n)=
-V_{qb}\frac{4G_F}{\sqrt{2}} \left[
{\cal A}_{ln}^q(\bar{q}\gamma^{\mu}P_Lb)(\bar{e}_l\gamma_{\mu}P_L\nu_n)-
{\cal B}_{ln}^q(\bar{q}P_Rb)(\bar{e}_lP_L\nu_n)\right],
\eeq
where we assume
the matrices of the soft mass terms are diagonal in the fermion mass basis.
Note that the operators in Eq. (4) take the same form as
those of the MSSM with $R_p$.  Comparing with the SM, the
above effective lagrangian includes the interactions even when $l$ and $n$ are
different from each other.
The dimensionless coupling constants
${\cal A}$ and ${\cal B}$ depend on the species of quark, charged lepton
and neutrino and are given by
\beqar
{\cal A}_{ln}^q&=&\frac{\sqrt{2}}{4G_FV_{qb}}\sum_{i,j=1}^{3}
\frac{1}{2m^2_{\tilde{d}_i^c}}V_{qj}\lambda'_{n3i}\lambda'^*_{lji},
\nonumber \\
{\cal B}_{ln}^q&=&\frac{\sqrt{2}}{4G_FV_{qb}}\sum_{i,j=1}^{3}
\frac{2}{m^2_{\tilde{l}_i}}V_{qj}\lambda_{inl}\lambda'^*_{ij3},
\eeqar
where $l$ and $n$ are the generation indices running from 1 to 3.

From the numerical values of \cite{PDG}, we find
\beqar
\ca_{ln}^u&=&\sum_{i=1}^{3}\lambda'_{n3i}
\left\{422\lambda'^*_{l1i}
~\left(\frac{V_{ud}/0.9751}{V_{ub}/0.0035}\right)
+96\lambda'^*_{l2i}
~\left(\frac{V_{us}/0.2215}{V_{ub}/0.0035}\right)
+1.52\lambda'^*_{l3i}\right\}
\left(\frac{100~ {\rm GeV}}{m_{\tilde{d}_i^c}}\right)^2,
\nonumber \\
\cb_{ln}^u&=&\sum_{i=1}^{3}\lambda_{inl}
\left\{1689\lambda'^*_{i13}
~\left(\frac{V_{ud}/0.9751}{V_{ub}/0.0035}\right)
+384\lambda'^*_{i23}
~\left(\frac{V_{us}/0.2215}{V_{ub}/0.0035}\right)
+6.1\lambda'^*_{i33}\right\}
\left(\frac{100~ {\rm GeV}}{m_{\tilde{l}_i}}\right)^2,
\nonumber \\
\ca_{ln}^c&=&\sum_{i=1}^{3}\lambda'_{n3i}
\left\{8.2\lambda'^*_{l1i}
~\left(\frac{V_{cd}/0.221}{V_{cb}/0.041}\right)
+36\lambda'^*_{l2i}
~\left(\frac{V_{cs}/0.9743}{V_{cb}/0.041}\right)
+1.52\lambda'^*_{l3i}\right\}
\left(\frac{100~ {\rm GeV}}{m_{\tilde{d}_i^c}}\right)^2,
\nonumber \\
\cb_{ln}^c&=&\sum_{i=1}^{3}\lambda_{inl}
\left\{32.7\lambda'^*_{i13}
~\left(\frac{V_{cd}/0.221}{V_{cb}/0.041}\right)
+144\lambda'^*_{i23}
~\left(\frac{V_{cs}/0.9743}{V_{cb}/0.041}\right)
+6.1\lambda'^*_{i33}\right\}
\left(\frac{100~ {\rm GeV}}{m_{\tilde{l}_i}}\right)^2.
\eeqar
Note the large numerical factors coming from the big differences between the
values of the CKM matrix elements.

First, we consider $q=u$ case.
At presents, the measurements of the branching ratios 
of the $B^\pm \ra l^\pm \nu$ 
processes give the upper bounds (at 90 $\%$ C.L.)\cite{PDG}
\beqar
BR( B^- \rightarrow  e^- \bar{\nu}_e) &<& 1.5 \times 10^{-5}, \nonumber \\
BR( B^- \rightarrow  \mu^- \bar{\nu}_\mu) &<& 2.1 \times 10^{-5}, \nonumber \\
BR( B^- \rightarrow  \tau^- \bar{\nu}_\tau) &<& 5.7 \times 10^{-4}. 
\eeqar
These experimental bounds are much larger than 
the standard model expectations;
$ BR( B^- \rightarrow  e^- \bar{\nu}_e )_{SM} \sim 9.2 \times 10^{-12}$,
$ BR( B^- \rightarrow  \mu^- \bar{\nu}_\mu )_{SM} \sim 3.9 \times 10^{-7}$
and
$ BR( B^- \rightarrow  \tau^- \bar{\nu}_\tau )_{SM} \sim 8.8 \times 10^{-5}$.
 
If we assume that the $R_p$-violating interactions are dominated, 
the decay rate of the processes $B^- \rightarrow e_l \bar{\nu}_n$  reads
\beqar
  \Gamma(B^- \rightarrow e_l \bar{\nu}_n) =
  {1 \over 8\pi} |V_{ub}|^2 G_F^2 f_B^2 M_B^3
  \bigg|{\cal A}^u_{ln} {m_l \over M_B} -{\cal B}^u_{ln} \bigg|^2 
  \left(1-{m_l^2 \over M_B^2}\right)^2,
\eeqar
using the PCAC (partial conservation of axial-vector current) relations
\beqar
\lag0|\bar{b}\gamma^{\mu}\gamma_5q|B_q(p)\rag &=&  i f_{B_q} p^{\mu}_{B_q},
\nonumber \\
\lag0|\bar{b}\gamma_5q|B_q(p)\rag &=& - i f_{B_q} \frac{M_{B_q}^2}{m_b+m_q}
\cong - i f_{B_q} M_{B_q},
\eeqar 

Since the species of the neutrinos cannot be distinguished by experiments
and the $R_p$-violating interactions allow the different kinds of the charged
lepton and the neutrino as decay products, we should sum the above decay
rates over neutrino species to compare with experimental data as follow
\beq
\Gamma (B^- \ra e^-_l \bar{\nu}) \equiv 
\sum_{n=1}^{3}\Gamma (B^- \ra e^-_l \bar{\nu}_n).
\eeq

%We then obtain
%\beqar
%  \bigg|A^u_{ln} {m_l \over M_B} -B^u_{ln} \bigg|^2 
% = 1020 
%    \left(0.2 {\rm GeV} \over f_B\right)^2
%    \left(5.28 {\rm GeV} \over M_B\right)^2
%    BR (B^- \rightarrow e^-_l \bar{\nu}_n),
%\eeqar

From the Eq. (8), (10) and the upper limit on the branching ratio, Eq. (7), 
we obtain
\beqar
\sum_{n=1}^{3} \bigg|{\cal B}^u_{1n} \bigg|^2 &<&  1.5 \times 10^{-2} ,\nl
\sum_{n=1}^{3} \bigg|{\cal B}^u_{2n} \bigg|^2 &<&  2.1 \times 10^{-2} ,\nl
\sum_{n=1}^{3} \bigg|{\cal B}^u_{3n}-0.337 {\cal A}^u_{3n} \bigg|^2 
&<&  7.4 \times 10^{-1} ,
\eeqar
For numerical calculations, we used $\tau_B=1.6~ ps$, $f_B=200~ MeV$,
$m_\tau = 1.78~ GeV$ and neglected the lepton masses for $l=e,\mu$ cases.

Under the assumption that only one product combination is not zero,
we get the upper bounds 
on some combinations of the $\lambda  {\lambda}^{\prime}$-
and ${\lambda}^{\prime} {\lambda}^{\prime}$-type. 
%\beqar
%|\lambda_{in1} \lambda'_{i13}| &<& 7.3 \times 10^{-5}, \nonumber\\  
%|\lambda_{in1} \lambda'_{i23}| &<& 3.2 \times 10^{-4}, \nonumber\\  
%|\lambda_{in1} \lambda'_{i33}| &<& 2.0 \times 10^{-2}, \nonumber\\  
%|\lambda_{in2} \lambda'_{i13}| &<& 8.7 \times 10^{-5}, \nonumber\\  
%|\lambda_{in2} \lambda'_{i23}| &<& 3.8 \times 10^{-4}, \nonumber\\  
%|\lambda_{in2} \lambda'_{i33}| &<& 2.4 \times 10^{-2}, \nonumber\\  
%|\lambda_{in3} \lambda'_{i13}| &<& 5.1 \times 10^{-4}, \nonumber\\
%|\lambda_{in3} \lambda'_{i23}| &<& 2.2 \times 10^{-3}, \nonumber\\
%|\lambda_{in3} \lambda'_{i33}| &<& 1.4 \times 10^{-1}, \nonumber\\
%|\lambda'_{n3i} \lambda'_{31i}| &<& 6.1 \times 10^{-3}, \nonumber\\
%|\lambda'_{n3i} \lambda'_{32i}| &<& 2.7 \times 10^{-2}, \nonumber\\
%|\lambda'_{n3i} \lambda'_{33i}| &<& 1.7 
%\eeqar
For the product combinations of $\lambda \lambda'$ type, 
we observe that the several bounds 
are stronger than the previous bounds and
list them in the Table I.
In the case of the product combinations of $\lambda' \lambda'$ type,
there is no stronger bound than the previous ones.
The previous bounds are calculated from the bounds on single $R_p$ violating
coupling, see Table 1 of ref. \cite{chahu}

Next, we consider $q=c$ case.
Recently charmed $B$ meson $(B_c)$ are observed \cite{bc}.
It is expected that in the near future large data sample of $B_c$ mesons
would be avaliable.
The standard model predictions of branching ratios for $B_c \ra l \nu$
decay modes are $BR (B_c \ra e \nu)_{SM} \sim 2.5 \times 10^{-9}$, 
$BR (B_c \ra \mu \nu)_{SM} \sim 1.0 \times 10^{-4}$, 
$BR (B_c \ra \tau \nu)_{SM} \sim 2.6 \times 10^{-2}$.
These branching ratios can be largely affected by R-parity violation.  
If we assume that the $R_p$-violating interactions are dominated, 
the decay rate of the processes $B_c^- \rightarrow e_l \bar{\nu}_n$  reads
\beqar
  \Gamma(B_c^- \rightarrow e_l \bar{\nu}_n) =
  {1 \over 8\pi} |V_{cb}|^2 G_F^2 f_{B_c}^2 M_{B_c}^3
  \bigg|{\cal A}^c_{ln} {m_l \over M_{B_c}} -{\cal B}^c_{ln} \bigg|^2 
  \left(1-{m_l^2 \over M_{B_c}^2}\right)^2.
\eeqar
In Table 2, we list the combinations of couplings whose present upper limits
allows the branching ratios to have the values of order of $10^{-2}$,
assuming only one product of $R_p$-violating couplings is nonzero.
For numerical calculations, we used $\tau_{B_c}=0.55~ ps$, $f_{B_c}=450~ MeV$,
$M_{B_c} = 6.275~ GeV$ ~\cite{mangano}.

%\vspace{1.0cm}
%\section*{conclusions}
%{\bf 3.}
To conclude, we have derived the more strigent 
upper bounds on certain products of $R_p$ and lepton-flavor-violating 
couplings from the upper limits of $B^\pm \ra l^\pm \nu$
branching ratios. 
And we investigate the possible effects of R-parity violating 
interactions on $B_c \ra l \nu$ decays.
These decay modes can be largely affected by R-parity violation. 

\section*{acknowledgements}
This work was supported in part by KOSEF postdoctoral program
(S.B) and the KAIST Center for Theoretical Physics and Chemistry (Y.G.K).

\begin{table}
\caption{\label{haha}
Upper bounds on the magnitudes of products of couplings
derived from $B \rightarrow l \nu$.
}
\begin{tabular}{llll}
%\hline
%%%%%%%%%%%%%%%%%%%%%%%%%%%%%%%%%%%%%%%%%%%%%%%%%%%%%%%%%%%%%
Decay Mode & Combinations Constrained & Upper bound & Previous bound \\
\hline
%%%%%%%%%%%%%%%%%%%%%%%%%%%%%%%%%%%%%%%%%%%%%%%%%%%%%%%%%%%%%
$B^- \rightarrow e^- \bar{\nu}$
%&$\llp{121}{123}$ & 3.2$\times 10^{-4}$ & 4.8 $\times 10^{-4}$ \\
%&$\llp{121}{223}$ &3.2$\times10^{-4}$  & 4.8 $\times 10^{-4}$ \\
&$\llp{131}{113}$ &7.3$\times10^{-5}$  & 4.9 $\times 10^{-4 a}$ \\
&$\llp{131}{123}$ &3.2$\times10^{-4}$  & 6.0 $\times 10^{-4}$ \\
&$\llp{131}{323}$ &3.2$\times10^{-4}$  & 6.0 $\times 10^{-4}$ \\
%&$\llp{131}{333}$ &2.0$\times10^{-2}$  & 2.6 $\times 10^{-2}$ \\
&$\llp{231}{213}$ &7.3$\times10^{-5}$  & 4.9 $\times 10^{-4 a}$ \\
&$\llp{231}{223}$ &3.2$\times10^{-4}$  & 5.5 $\times 10^{-4}$ \\
&$\llp{231}{233}$ &2.0$\times10^{-2}$  & 2.0 $\times 10^{-2}$ \\
&$\llp{231}{323}$ &3.2$\times10^{-4}$  & 5.5 $\times 10^{-4}$ \\
%&$\llp{231}{333}$ &2.0$\times10^{-2}$  & 2.3 $\times 10^{-2}$ \\
\hline
%%%%%%%%%%%%%%%%%%%%%%%%%%%%%%%%%%%%%%%%%%%%%%%%%%%%%%%%%%%%%
$B^- \rightarrow \mu^- \bar{\nu}$
%&$\llp{122}{223}$ &3.8$\times10^{-4}$  & 4.8 $\times 10^{-4}$ \\
&$\llp{132}{113}$ &8.7$\times10^{-5}$  & 6.0 $\times 10^{-4 a}$ \\
&$\llp{132}{123}$ &3.8$\times10^{-4}$  & 6.0 $\times 10^{-4}$ \\
&$\llp{132}{323}$ &3.8$\times10^{-4}$  & 6.0 $\times 10^{-4}$ \\
%&$\llp{132}{333}$ &2.4$\times10^{-2}$  & 2.6 $\times 10^{-2}$ \\
&$\llp{232}{213}$ &8.7$\times10^{-5}$  & 6.0 $\times 10^{-4 a}$ \\
&$\llp{232}{223}$ &3.8$\times10^{-4}$  & 5.5 $\times 10^{-4}$ \\
%&$\llp{232}{233}$ &2.4$\times10^{-2}$  & 4.0 $\times 10^{-2}$ \\
\hline
%%%%%%%%%%%%%%%%%%%%%%%%%%%%%%%%%%%%%%%%%%%%%%%%%%%%%%%%%%%%%
$B^- \rightarrow \tau^- \bar{\nu}$
&$\llp{123}{113}$ &5.1$\times10^{-4}$  & 6.0 $\times 10^{-4 a}$ \\
&$\llp{233}{213}$ &5.1$\times10^{-4}$  & 5.5 $\times 10^{-4}$ \\
&$\llp{233}{313}$ &5.1$\times10^{-4}$  & 6.0 $\times 10^{-4 a}$ \\
%%%%%%%%%%%%%%%%%%%%%%%%%%%%%%%%%%%%%%%%%%%%%%%%%%%%%%%%%%%%%
\end{tabular}
a:  Bounds from $B \rightarrow \l^+_i \l^-_j$\cite{Lee}.
%Others : Table 1 of \cite{chahu}.
\end{table}

\begin{table}
\caption{\label{hahaha}
Maximally allowed braching ratios and the list of combinations
whose present upper bounds allow the branching ratios to have
the value of order of $10^{-2}$. 
}
\begin{tabular}{llll}
%\hline
%%%%%%%%%%%%%%%%%%%%%%%%%%%%%%%%%%%%%%%%%%%%%%%%%%%%%%%%%%%%%
Decay Mode & Combinations & Branching Ratio \\
% & Constrained &  &  \\
\hline
%%%%%%%%%%%%%%%%%%%%%%%%%%%%%%%%%%%%%%%%%%%%%%%%%%%%%%%%%%%%%
$B_c^- \rightarrow e^- \bar{\nu}$&
$\llp{131}{123}$ & 1.1$\times 10^{-2}$ \\
&$\llp{131}{323}$ &1.1$\times10^{-2}$  \\
&$\llp{231}{223}$ &2.2$\times10^{-2}$  \\
\hline
%%%%%%%%%%%%%%%%%%%%%%%%%%%%%%%%%%%%%%%%%%%%%%%%%%%%%%%%%%%%%
$B_c^- \rightarrow \mu^- \bar{\nu}$
&$\llp{132}{123}$ &1.0$\times10^{-2}$  \\
&$\llp{232}{223}$ &2.1$\times10^{-2}$  \\
\hline
%%%%%%%%%%%%%%%%%%%%%%%%%%%%%%%%%%%%%%%%%%%%%%%%%%%%%%%%%%%%%
$B_c^- \rightarrow \tau^- \bar{\nu}$
&$\llp{233}{223}$ &0.5$\times10^{-2}$  \\
%%%%%%%%%%%%%%%%%%%%%%%%%%%%%%%%%%%%%%%%%%%%%%%%%%%%%%%%%%%%%
\end{tabular}
\end{table}

\end{document}